\documentclass[a4paper]{article}

\usepackage[margin=2.5cm]{geometry}
\usepackage{algorithm}
\usepackage{algpseudocode}
\usepackage{amssymb}
\usepackage{amsmath}
\usepackage{hyperref}
\usepackage{cleveref}
\usepackage{autonum}
\usepackage{subcaption}
\usepackage{pgfplots}
\usepackage{authblk}

\pgfplotsset{compat=1.3, cycle list name=exotic}

\title{2D implementation of Kinetic-diffusion Monte Carlo in Eiron}
\author[1]{Oskar Lappi\footnote{Equal contribution}}
\newcommand\CoAuthorMark{\footnotemark[\arabic{footnote}]}
\author[2]{Emil Løvbak\protect\CoAuthorMark}
\author[3]{Thijs Steel}
\author[3]{Giovanni Samaey}

\date{}

\affil[1]{Department of Computer Science, University of Helsinki, Helsinki, Finland}
\affil[2]{Scientific Computing Center, Karlsruhe Institute of Technology, Karlsruhe, Germany}
\affil[3]{Department of Computer Science, KU Leuven, Leuven, Belgium}

\begin{document}
	
\maketitle

\begin{abstract}
Particle-based kinetic Monte Carlo simulations of neutral particles is one of the major computational bottlenecks in tokamak scrape-off layer simulations. This computational cost comes from the need to resolve individual collision events in high-collisional regimes. However, in such regimes, one can approximate the high-collisional kinetic dynamics with computationally cheaper diffusion. Asymptotic-preserving schemes make use of this limit to perform simulations in these regimes, without a blow-up in computational cost as incurred by standard kinetic approaches. One such scheme is Kinetic-diffusion Monte Carlo. In this paper, we present a first extension of this scheme to the two-dimensional setting and its implementation in the Eiron particle code. We then demonstrate that this implementation produces a significant speedup over kinetic simulations in high-collisional cases.
\end{abstract}

\section{Introduction}
\label{sec:intro}

Particle-based, kinetic Monte Carlo simulation is a key component in modeling of neutral particles in the tokamak scrape-off layer~\cite{Borodin2022}. The EIRENE code~\cite{Reiter2005} is widely used to perform such simulations, as part of, e.g., the SOLPS-ITER software~\cite{Wiesen2015}; however, we note the DEGAS code~\cite{Stotler1994} as a widely used alternative. A drawback of kinetic Monte Carlo simulation is that it requires the resolution of individual collision events at the particle level. Resolving all of these collisions can become computationally infeasible, especially in high-collisional regimes. Mathematically, it is known that one can approximate kinetic models by drift-diffusion type models in this regime, assuming a diffusive scaling~\cite{Markowich1990}, thus avoiding the need to resolve all collisions.

A variety of hybrid approaches exist that combine kinetic simulation with diffusive simulation~\cite{Borodin2022}, including approaches that combine Monte Carlo simulation with deterministic methods, e.g.~\cite{Blommaert2019, Horsten2020}. However, we make use of asymptotic-preserving Monte Carlo schemes (APMC)~\cite{Mortier2022, Dimarco2018}, which work fully at the Monte Carlo level. Specifically, we make use of the kinetic-diffusion Monte Carlo method (KDMC)~\cite{Mortier2022}. We note that these schemes are a special case of a broader field of asymptotic-preserving schemes, see~\cite{Jin2022} and references therein.

The general idea behind APMC schemes is that they simulate particle trajectories through time-stepping that alternates between kinetic and diffusive positional increments in such a way that mainly simulate kinetic dynamics in low-collisional (kinetic) regimes, and diffusive dynamics in high-collisional (diffusive) regimes. We note that these schemes, in general, do not provide any error guarantees in the intermediate regimes. Any errors in these regimes can however be removed by the combination of APMC with multilevel Monte Carlo techniques~\cite{Giles2008, Heinrich2001}, see~\cite{Mortier2022a, Loevbak2021, Loevbak2023}. In the cited works, the APMC approach is shown to significantly reduce the required computational effort for simulating simplified mathematical test problems.

Implementing APMC methods in EIRENE is a significant undertaking, due to the code's complexity and scope. Hence, we assess their feasibility in a simpler code, with limited physics, namely the Eiron code~\cite{Lappi2025}. Eiron is a newly developed \verb|C++| code, following modern software design principles. The code serves as a sandbox in which to do performance analysis and test new numerical and programming approaches at low development effort, before embarking on time intensive implementation in state-of-the-art production codes.

In this paper, we present the first two-dimensional implementation of KDMC. We have performed this implementation in Eiron, allowing us to make use of it's efficient, modular implementation to perform HPC-scale simulations. We present the technical challenges of implementing this mathematical algorithm in an efficient and robust way and perform an initial analysis of the resulting simulation accuracy and computational speedup. The remainder of this paper is structured as follows. In Section~\ref{sec:kdmc}, we provide an introduction to and algorithmic description of the KDMC method. Next, in Section~\ref{sec:eiron}, we discuss relevant practical aspects of the Eiron code and the details of the KDMC implementation. In Section~\ref{sec:results}, we then present some simulation results and compare the convergence behavior of the Eiron implementation to the original 1D results in~\cite{Mortier2022}. We also conduct a performance analysis demonstrating the computational advantage of KDMC in high-collisional simulations. Finally, in Section~\ref{sec:conclusions}, we draw conclusions from the presented work and comment on future work.

\section{Monte Carlo for neutral particle models}
\label{sec:kdmc}

We consider the simulation of neutral particles, based on a Boltzmann-BGK~\cite{Bhatnagar1954} equation
\begin{equation}
	\label{eq:kinetic_equation}
	\partial_t \, f(x,v,t) + v \cdot \nabla_x \, f(x,v,t) = R_\text{cx} \left( \rho(x,t)\mathcal{M}(v) - f(x,v,t) \right).
\end{equation}
Here, $f(x,v,t)$ represents the density of particles in a phase space over position $x$ and velocity $v$, as a function of time $t$. On the right-hand side of \eqref{eq:kinetic_equation}, we model charge exchange collisions with a homogeneous event rate $R_\text{cx}$. In the BGK collision operator, we make use of $\rho(x,t)$ the particle density (integrating $f(x,v,t)$ over velocity space) and $\mathcal{M}(v)$ a Maxwellian distribution, with a given temperature $T$ and mean drift velocity $u$. Although we constrain ourselves to homogeneous test-cases, both for the sake of exposition and for performing the initial experiments presented in this work, we note that neither Eiron, nor the KDMC approach are inherently constrained to such cases.

\subsection{Kinetic Monte Carlo}

Neglecting domain boundary interactions, we can perform a kinetic Monte Carlo simulation of \eqref{eq:kinetic_equation} by simulating random particle trajectories with collision events samples according to the Maxwellian $\mathcal{M}(v)$. We present an algorithmic description of such a trajectory simulation in Algorithm~\ref{alg:kmc}. Note here that $\mathcal{E}(R_\text{cx})$ denotes an exponential distribution with rate $R_\text{cx}$. Due to resolving all collision events, kinetic Monte Carlo becomes prohibitively expensive in highly collisional regimes.

\begin{algorithm}
	\begin{algorithmic}[1]
		\State{$t\gets 0, \quad n\gets 0, \quad \left\{X_p^0, V_p^0\right\} \gets \text{sample\_source()}$}
		\While{$t<T$}
		\State{$\tau \sim \mathcal{E}\left(R_\text{cx}\right)$}
		\State{$X_p^{n+1} \gets X_p^n +\tau V^n_p$}
		\State{$V^{n+1}_p \sim \mathcal{M}(v)$}
		\State{$n \gets n+1, \quad t \gets t + \tau$}
		\EndWhile
	\end{algorithmic}
	\caption{Simulating a particle with kinetic Monte Carlo until time $T$.\label{alg:kmc}}
\end{algorithm}

\subsection{Kinetic-diffusion Monte Carlo}

The kinetic-diffusion Monte Carlo method (KDMC) approximates highly-collisional kinetic Monte Carlo simulations by introducing diffusion at the particle level. The algorithm alternates between kinetic and diffusive updates to a given particle's position. The general idea behind this approach is that, in the infinite collisional limit, one replaces an infinite number of kinetic positional updates (Algorithm~\ref{alg:kmc}, line 4) with a single normally distributed increment, through the application of the law of large numbers. We refer to~\cite{Mortier2022} for a detailed description of the algorithm.

We generalize the derivations in~\cite{Mortier2020} to the two-dimensional setting, to observe that a correct implementation requires computing diffusive increments $\Delta W_p^n$ that are sampled from a multivariate normal distribution $\mathcal{N}(\mu, \Sigma)$, with mean
\begin{equation}
	\mu = u\theta + \frac{V_p^{n+1} - u}{R_\text{cx}}\left( 1 - e^{-\theta R_\text{cx}} \right),
	\label{eq:kdmean}
\end{equation}
and covariance matrix
\begin{equation}
	\Sigma = \frac{2T}{R_\text{cx}^2\theta}\left( 2e^{-\theta R_\text{cx}} + \theta R_\text{cx}\left(1 + e^{-\theta R_\text{cx}} \right) - 2 \right) I + \left( 1 - 2\theta R_\text{cx} e^{-\theta R_\text{cx}} - e^{-2\theta R_\text{cx}} \right) \left( \frac{V_p^{n+1} - u}{R_\text{cx}} \right)^{\otimes2},
	\label{eq:kdvariance}
\end{equation}
where the exponent $\otimes 2$ is used to denote the outer product of a vector with itself. Here, $V_p^{n+1}$ is the velocity of the subsequent kinetic increment, and $\theta$ is a diffusive flight time. Algorithm~\ref{alg:kdmc} shows a KDMC simulation of a single trajectory until time $T$.

\begin{algorithm}
	\begin{algorithmic}[1]
		\State{$t\gets 0, \quad n \gets 0, \quad \left\{X_p^0, V_p^0\right\} \gets \text{sample\_source()}$}
		\While{$t<T$}
		\State{$\tau \sim \mathcal{E}\left(R_\text{cx}\right)$}
		\State{$X_p^{n,\prime} \gets X_p^n +\tau V^n_p$}
		\State{$V^{n+1}_p \sim \mathcal{M}(v)$}
		\State{$\theta \gets \Delta t - (\tau \; \text{mod} \; \Delta t)$}
		\State{$\Delta W_p^n \sim \mathcal{N}(\mu, \Sigma)$
			\State{$X_p^{n+1} \gets X_p^{n,\prime} + \Delta W$}}		
		\State{$n \gets n+1, \quad t \gets t + \Delta t$}
		\EndWhile
	\end{algorithmic}
	\caption{Simulating a particle with KDMC until time $T$.\label{alg:kdmc}}
\end{algorithm}

We refer to~\cite{Mortier2022} for further details and a full analysis of the scheme, but make the following observations in terms of the charge-exchange rate $R_\text{cx}$ in relation to the KDMC discretization time step size $\Delta t$. In low-collisional regimes, i.e., $R_\text{cx} \ll \Delta t^{-1}$, observe that, in general, $\tau \gg \theta$. Hence, the kinetic component (line 4) of Algorithm~\ref{alg:kdmc} will dominate and the results produced by Algorithm~\ref{alg:kdmc} will converge to those produced by Algorithm~\ref{alg:kmc}. Inversely, in high-collisional regimes, i.e., $R_\text{cx} \gg \Delta t^{-1}$, we have that, in general, $\tau \ll \theta$. Hence, the diffusive component (line 8) of Algorithm~\ref{alg:kdmc} will dominate. An intuitive explanation for the introduction of the diffusive step in line 8 is that one substitutes the sum of an infinite number of i.i.d. steps of the form given in line 4 by a single normally distributed increment, making use of the law of large numbers.

\section{Implementation in Eiron}
\label{sec:eiron}

Eiron is a reduced 2D kinetic Monte Carlo simulation code, created to study the performance of the different Monte Carlo parallelization strategies that could be used in EIRENE.
We have implemented the KDMC simulation algorithm (Algorithm~\ref{alg:kdmc}) in Eiron, reusing existing functionality for kinetic simulations. Hence, only the diffusive step required implementation from scratch. We are also able to use the existing parallel algorithms in Eiron to run KDMC simulations, as these are decoupled from the simulation and estimation code.

Eiron supports a variety of shared and distributed-memory parallelism.
Ideally, all parallelization strategies would be available to KDMC; unfortunately KDMC breaks some assumptions in Eiron's domain-decomposition algorithm, hence constraining us to a single domain.
Unlike kinetic transport, a diffusive transport step is not just a continuous integral that can easily be broken at a subdomain boundary and continued in a neighboring subdomain boundary.
When simulating diffusive steps for heterogeneous backgrounds, we approximately evaluate $R_\text{cx}$ at the midpoint of the step using an estimated endpoint computed with $\mu$ evaluated at $X^n_p$ and $\Sigma=0$. If this estimated endpoint lies in a separate subdomain, this procedure introduces additional communication, not currently supported by Eiron.
In future work, we plan to resolve these technical issues and implement domain decomposition for KDMC.

\section{Simulation results}
\label{sec:results}

We now present the performance of our implementation, both in terms of accuracy and computational efficiency.
In Section~\ref{sec:resultskinetic}, we consider the low-collisional, kinetic test case ($R_\text{cx} \ll \Delta t^{-1}$).
Here we show that KDMC and kinetic simulations converge to the same results as $\Delta t \to 0$, with comparable computational cost.
In Section~\ref{sec:resultsdiffusive}, we consider a high-collisional, diffusive test case ($R_\text{cx} \gg \Delta t^{-1}$).
Here we show that KDMC and kinetic simulations also converge as $R_\text{cx} \to \infty$, with the KDMC simulations significantly outperforming the kinetic simulations in terms of computational cost.
In general, one expects the largest errors to occur when $\Delta t$ and $R_\text{cx}$ are of the same order of magnitude.

Throughout this Section, our test-case is a two-dimensional slab with ionizing boundaries. In what follows, we normalize all simulation parameters, assuming that the domain has dimensions $1\;\text{m} \times 1\;\text{m}$. We initialize all particles using an isotropic point-source at the center of the domain with velocities sampled from a 2D-Maxwellian. We then simulate the particles to a given end time $t_\text{end}$, at which point we construct a scaled histogram of their final location in the domain using a $128 \times 128$-grid, such that the solution integrates to one. We consider homogeneous collision backgrounds, with a fixed charge-exchange rate $R_\text{cx}$. Post-collisional velocities are sampled from an unbiased isotropic Maxwellian. We elaborate on precise numerical parameter values when discussing individual test-cases.

The simulations discussed in this section were run on CSC's Mahti supercomputer~\cite{CSC2025} using 128 OpenMP threads all running on separate CPU cores.

\subsection{Kinetic regime}
\label{sec:resultskinetic}

\begin{figure}
	\centering
	\begin{subfigure}{0.45\textwidth}
                \includegraphics[width=\textwidth]{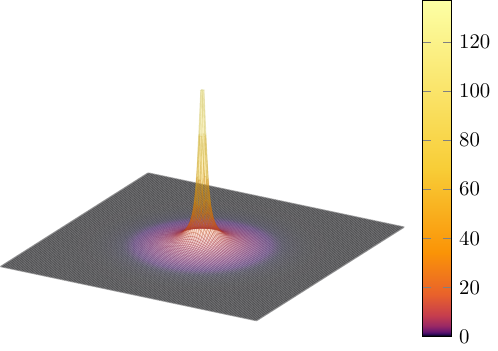}
		\centering
                \caption{KDMC result}
	\end{subfigure}
	\hfill
	\begin{subfigure}{0.45\textwidth}
                \includegraphics[width=\textwidth]{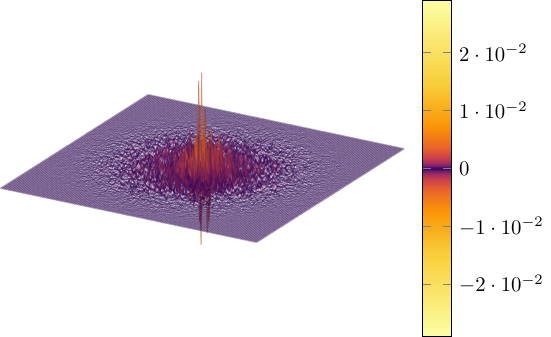}
		\centering
                \caption{Difference with kinetic simulation}
	\end{subfigure}
	\caption{A two-dimensional histogram of a KDMC simulation in the kinetic regime with $\Delta t=2^{-4}\;\text{s}$, $R_\text{cx}=0.78125 \;\text{s}^{-1}$ and a mean post-collisional speed of $0.013847\; \text{ms}^{-1}$.  To compute the pointwise difference we subtract the KDMC result from the kinetic result.}\label{fig:kinetic2D}
\end{figure}

When refining the time step parameter $\Delta t$, KDMC should converge to a kinetic simulation. In the 1D setting, it has been shown that both particle distributions converges $\mathcal{O}(\Delta t^{1.5})$ in Wasserstein-1 distance for $\Delta t \to 0$~\cite{Mortier2022}. However, we note that this theoretical bound cannot be applied directly to our setting. To verify this limit we set a source mean speed of $0.15625 \;\text{ms}^{-1}$, the charge-exchange rate is $R_\text{cx}=0.78125$ and the post-collisional velocities have mean speed $0.013847 \;\text{ms}^{-1}$. We then run both kinetic and KDMC simulations with $10^{10}$ particles until $t_\text{end}=1\;\text{s}$ for a sequence of time step values $\Delta t$ in the range $[2^{-4}\;\text{s}, 1\;\text{s}]$.

\begin{figure}
	\centering
	\begin{subfigure}{0.45\textwidth}
		\centering
		\resizebox{0.8\textwidth}{!}{
			\begin{tikzpicture}[trim axis left]
				\begin{axis}[ymode=log, xmode=log, xlabel={$\Delta t$ [s]}, ylabel={Error 2-norm}, ylabel near ticks, x dir=reverse]
					\addplot table [x=delta_t, y=error, col sep=comma] {data_archive/clean/kinetic_limit_10B/kinetic_10B_convergence.csv};			
				\end{axis}
		\end{tikzpicture}}
		\caption{2-norm of the differences as a function of $\Delta t$.}\label{fig:kineticconvergenceplot}
	\end{subfigure}\hspace{0.05\textwidth}
	\begin{subfigure}{0.45\textwidth}
		\centering
		\resizebox{0.8\textwidth}{!}{
			\begin{tikzpicture}[trim axis left]
			\begin{axis}[xmode=log, xlabel={$\Delta t$ [s]}, ylabel={Runtime [s]}, ylabel near ticks, legend pos=north west, ymin=0, ymax=300, x dir=reverse]
				\addplot table [x=delta_t, y=time_kinetic, col sep=comma] {data_archive/clean/kinetic_limit_10B/kinetic_10B_runtime.csv};
				\addlegendentry{Kinetic}
				\addplot table [x=delta_t, y=time_kdmc, col sep=comma] {data_archive/clean/kinetic_limit_10B/kinetic_10B_runtime.csv};
				\addlegendentry{KDMC}				
			\end{axis}
			\end{tikzpicture}}
		\caption{Runtime as a function of $\Delta t$.}\label{fig:kinetictiming}
	\end{subfigure}
	\caption{Behavior of KDMC, compared to kinetic simulation, in the kinetic regime. To reduce variance, we average the 2D results in the $x$-direction and fold the the results around the origin.}\label{fig:kineticconvergence}
\end{figure}
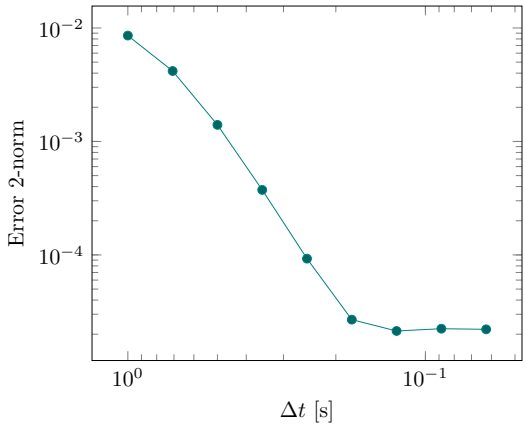
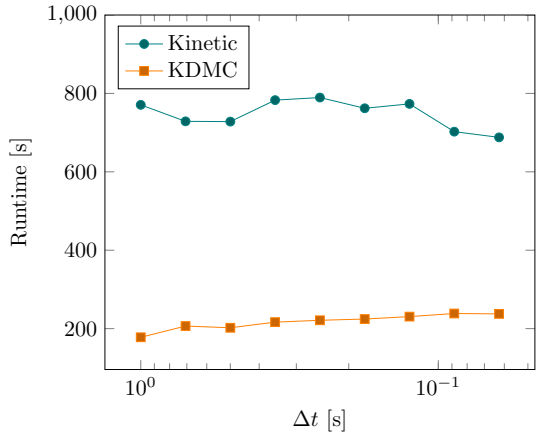

To visualize the results, we show a KDMC simulation for $\Delta t=2^{-4}\;\text{s}$ in Figure~\ref{fig:kinetic2D}, together with a cell-by-cell difference with the reference kinetic simulation. As the two simulations are visually indistinguishable, we opt not to show the reference separately. Comparing the computed solution with the difference, we observe that we achieve between three and four digits of accuracy in the region of interest at the center of the domain. 

In Figure~\ref{fig:kineticconvergence}, we consider the convergence of the KDMC simulation. To reduce variance, we flatten the solution to one dimension, by averaging over the x-dimension. We then mirror the solution around the source at the middle of the domain. In Figure~\ref{fig:kineticconvergenceplot}, we plot the 2-norm of the differences of these curves. We observe that this two norm decreases with roughly an order 3 until it hits a plateau caused by the sampling error due to the finite number of trajectories. This indicates a convergence that is roughly one and a half orders better than that predicted by the theoretical results in~\cite{Mortier2022}. Next to the fact that there is no direct link between the Wasserstein metric on the particles' location and the 2-norm on the histogram, it is likely that the symmetry of the problem is a cause for this increased rate of convergence. In Figure~\ref{fig:kinetictiming}, we show the corresponding run times and observe that the KDMC and kinetic simulations incur a comparable computational cost in the kinetic regime.

\subsection{Diffusive regime}
\label{sec:resultsdiffusive}

When increasing the collision rate, the modeling error incurred by approximating kinetic increments by diffusion in the KDMC scheme should vanish. In~\cite{Mortier2022}, it was show that this convergence should occur as $\mathcal{O}(R_\text{cx}^{-1.5})$, in terms of Wasserstein-1 distance of the particle distribution, as $R_\text{cx}\to\infty$. To verify the limit we perform simulations with a similar particle source with mean speed $0.0625 \;\text{ms}^{-1}$. We fix $\Delta t = 1\;\text{s}$ and consider combinations the mean post-collisional speed $\frac{1}{\varepsilon} \sqrt{\frac{\pi}{10}}\frac{1}{512}\;\text{ms}^{-1}$ and collision rate $R_\text{cx}=\frac{1}{\varepsilon^2}\frac{1}{128}\;\text{s}^{-1}$ with $\varepsilon$ taking values in the range $[2^{-7.5},1]$. In this case we use $2\cdot 10^9$ particles until time $t_\text{end}=4\;\text{s}$. We then generate similar figures to those in Section~~\ref{sec:resultskinetic}.

\begin{figure}
	\centering
	\begin{subfigure}{0.45\textwidth}
                \includegraphics[width=\textwidth]{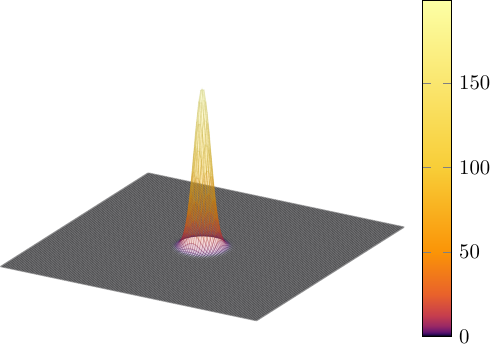}
		\centering
                \caption{KDMC result}
	\end{subfigure}
	\hfill
	\begin{subfigure}{0.45\textwidth}
                \includegraphics[width=\textwidth]{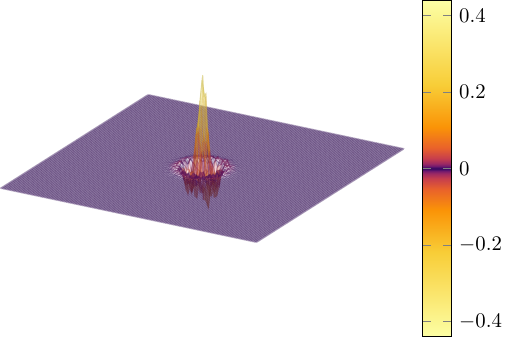}
		\centering
		\caption{Difference with kinetic simulation}
	\end{subfigure}
	\caption{A two-dimensional histogram of a KDMC simulation in the diffusive regime with $\Delta t=1\;\text{s}$, $R_\text{cx}=256 \;\text{s}^{-1}\vphantom{\text{s}^{-1}}$ and a mean post-collisional speed of $0.19817\;\text{ms}^{-1}$.  To compute the pointwise difference we subtract the KDMC result from the kinetic result.}\label{fig:diffusive2D}
\end{figure}

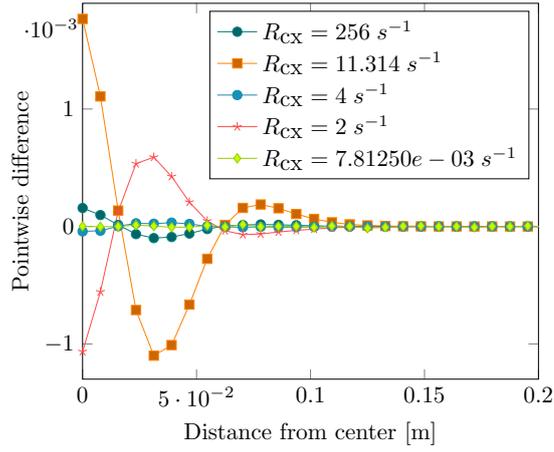
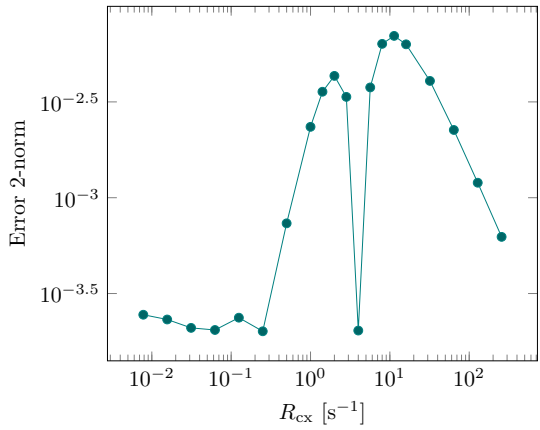
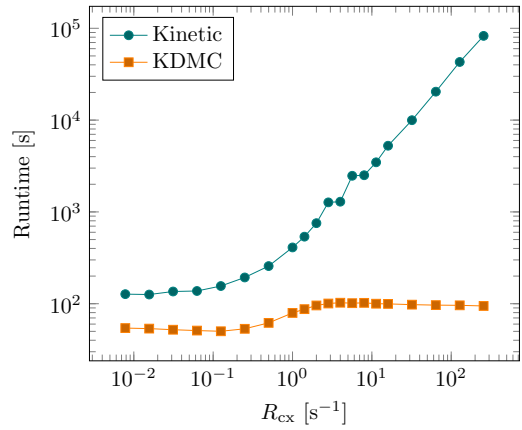
\begin{figure}
	\centering
	\begin{subfigure}{0.5\textwidth}
		\centering
		\resizebox{0.8\textwidth}{!}{
			\begin{tikzpicture}[trim axis left]
			\begin{axis}[
				ymax=2.4e-1,
				ymin=-1.7e-1,
				xmax=0.2,
				xmin=0,
				xlabel={Distance from center [m]},
				ylabel={Pointwise difference}, ylabel near ticks,
				y tick scale label style={yshift=-0.7cm, xshift=-1cm},
				legend cell align=left]
				
				\foreach \val in {256,11.314,4,2,7.81250e-03} {
					\addplot table [x=x, y=\val, col sep=comma] {data_archive/clean/diffusive_limit_2B/diffusive_2B_reduced_x_folded_profiles_wrapped.csv} ;
					\edef\temp{\noexpand\addlegendentry{$R_{\text{cx}}=\val \; s^{-1}$}}
					\temp
				}
			\end{axis}
			\end{tikzpicture}}
		\caption{Pointwise difference for a selection of values of $R_{\text{cx}}$}\label{fig:diffusiveoverlay}
	\end{subfigure}\vspace{0.2cm}
	
	\begin{subfigure}{0.45\textwidth}
		\centering
		\resizebox{0.8\textwidth}{!}{
			\begin{tikzpicture}[trim axis left]
				\begin{axis}[ymode=log, xmode=log, xlabel={$R_\text{cx}$ [$\text{s}^{-1}$]}, ylabel={Error 2-norm}, ylabel near ticks]
					\addplot table [x=Rcx, y=error, col sep=comma] {data_archive/clean/diffusive_limit_2B/diffusive_2B_convergence_Rcx.csv};			
				\end{axis}
		\end{tikzpicture}}
		\caption{2-norm of the differences as a function of $R_\text{cx}$}\label{fig:diffusiveconvergenceplot}
	\end{subfigure}%
		\hspace{0.05\textwidth}%
		\begin{subfigure}{0.45\textwidth}
		\centering
		\resizebox{0.8\textwidth}{!}{
	\begin{tikzpicture}[trim axis left]
		\begin{axis}[ymode=log, xmode=log, xlabel={$R_\text{cx}$ [$\text{s}^{-1}$]}, ylabel={Runtime [s]}, ylabel near ticks, legend pos=north west]
			\addplot table [x=Rcx, y=time_kinetic, col sep=comma] {data_archive/clean/diffusive_limit_2B/diffusive_2B_runtime_Rcx.csv};
			\addlegendentry{Kinetic}
			\addplot table [x=Rcx, y=time_kdmc, col sep=comma] {data_archive/clean/diffusive_limit_2B/diffusive_2B_runtime_Rcx.csv};
			\addlegendentry{KDMC}				
		\end{axis}
		\end{tikzpicture}}
		\caption{Runtime as a function of $R_\text{cx}$}\label{fig:diffusivetiming}
	\end{subfigure}
	\caption{Behavior of KDMC, compared to kinetic simulation, in the diffusive regime. To reduce variance, we average the 2D results in the $x$-direction and fold the the results around the origin.}\label{fig:diffusiveconvergence}
\end{figure}

In Figure~\ref{fig:diffusive2D}, we show the KDMC simulation and pointwise error for the case $R_\text{cx}=256 \;\text{s}^{-1}$ and the mean post-collisional speed $0.19817\;\text{ms}^{-1}$. Note that the shape of the solution differs from that for the kinetic case, in that it resembles a bell-curve, as opposed to the exponential decay observed in Figure~\ref{fig:kinetic2D}, which was faster closer to the center source but slower further away. 
We consider the convergence behavior in Figure~\ref{fig:diffusiveconvergence}. In Figure~\ref{fig:diffusiveconvergenceplot}, we empirically observe a convergence rate of with approximate order 1 for $R_\text{cx} \geq 16\;\text{s}^{-1}$. Notably, this convergence is slower than the known bound in the Wasserstein-1 distance. However, in parallel work~\cite{Tang2025} a similar rate was shown in the $L_2$-norm, when computing time-integrated quantities using KDMC, combined with the diffusive estimator presented in~\cite{Mortier2022b}.

For $R_\text{cx} \leq 2\;\text{s}^{-1}$, we see a similar phenomenon in Figure~\ref{fig:diffusiveconvergenceplot} to that shown in Figure~\ref{fig:kineticconvergenceplot}, i.e., as $R_\text{cx}$ decreases, the both simulations converge to kinetic simulations until a plateau is achieved due to the sampling error. In the intermediate regime, the relative 2-norm of the differences achieves a maximum of less than 1\%. What is initially surprising, is that we observe a sharp dip approximately $R_\text{cx}=4\;\text{s}^{-1}$. To explain this phenomenon, we plot the pointwise differences of the averaged and folded simulation results for a selection of values of $R_\text{cx}$ in Figure~\ref{fig:diffusiveoverlay}.
In this figure, we see that the error curves for the averaged and flattened results change sign at one point (compare the curves for $R_\text{cx}=2\;\text{s}^{-1}$ [peak error on the kinetic side of Figure~\ref{fig:diffusiveconvergenceplot}] and $R_\text{cx}=11.314\;\text{s}^{-1}$ [peak error on the diffusive side of Figure~\ref{fig:diffusiveconvergenceplot}]).
At the point of transition, where the sign of the error changes  (roughly $R_\text{cx}=4\;\text{s}^{-1}$), there is therefore an artificially reduced error.

In Figure~\ref{fig:diffusivetiming}, we show that KDMC achieves a significant speedup in the diffusive regime, growing proportionally to $R_\text{cx}$. Here it is clear that the runtime of the kinetic simulations directly scale with the number of collision events that need to be resolved, while that for the KDMC simulations stays more-or-less constant. In the highest-collisional cases, this results in a speedup of two orders of magnitude, showing the strength of KDMC over standard kinetic simulation in this regime.

\section{Conclusions}
\label{sec:conclusions}

We have introduced the first two-dimensional implementation of the KDMC scheme for the  Boltzman-BGK equation in the Eiron code. Using this implementation, we confirmed that the results produced by KDMC converge to those produced by a standard kinetic simulation in both the kinetic limit ($\Delta t \to 0$) and diffusive limit ($R_\text{cx} \to \infty$). Our measured 2-norm convergence, does not match the rates given in~\cite{Mortier2022} for the Wasserstein-1 distance. However, we believe the improved rate in $\Delta t$ to be due to the symmetry in the considered test-case and our rate in the limit $R_\text{cx} \to 0$ matches the $L_2$-norm results shown in parallel work on analyzing time-integrated simulations with KDMC. The maximum observed relative error on the averaged and folded results is less that 1\% in the considered test cases. Care should be taken in generalizing this observation beyond these test cases as they do not contain, e.g., time averaging source-term estimation procedures such as those used in practical scrape-off layer simulations. However, we note that multilevel Monte Carlo approaches inspired by~\cite{Mortier2022a,Loevbak2021,Loevbak2023} can be leveraged for further error reduction in these regions if needed. In the kinetic regime, KDMC simulations attain similar run times to the equivalent kinetic simulations. However, in diffusive regimes, they outperform the kinetic reference simulations. In some cases, by multiple orders of magnitude.

This implementation in Eiron, is a first step towards the use of KDMC for large scale fusion simulations. In future work, we plan to extend the implementation to make full use of the domain-decomposition available in Eiron. We also plan to tackle more generic classes of boundary conditions such as reflective boundaries, in addition to improving the existing (biased) implementation of absorbing boundaries. One strategy for this could be the work presented in~\cite{Steel2025}. We also plan to run more extensive experiments for non-homogeneous test-cases. Further down the road, we plan to support time-integrated simulations by implementing suitable estimators for the diffusive increments, e.g., the scheme presented in~\cite{Mortier2022b}.

\section*{Author contributions}

\textbf{Oskar Lappi/Emil Løvbak:} Methodology, Software, Validation, Investigation, Writing - Original draft, Visualization \\ \textbf{Thijs Steel:} Methodology, Validation, Investigation, Writing - Review \& Editing \\ \textbf{Giovanni Samaey:} Conceptualization, Supervision, Writing - Review \& Editing, Funding acquisition

\section*{Acknowledgments}

We thank the anonymous reviewers for their efforts in improving this manuscript. The authors wish to acknowledge CSC – IT Center for Science, Finland, for computational resources.
Emil Løvbak was funded by the Deutsche Forschungsgemeinschaft (DFG, German Research Foundation) – Project-ID 563450842. This work has been carried out within the framework of the EUROfusion Consortium, funded by the European Union via the Euratom Research and Training Programme (Grant Agreement No 101052200 --- EUROfusion). Views and opinions expressed are however those of the author(s) only and do not necessarily reflect those of the European Union or the European Commission. Neither the European Union nor the European Commission can be held responsible for them. 

\section*{Data availability}

A copy of the data used to generate all figures, together with documentation on its generation can be found at \href{https://doi.org/10.23729/fd-08ac1189-361d-3350-a66f-99e9db473b30}{https://doi.org/10.23729/fd-08ac1189-361d-3350-a66f-99e9db473b30}.

\bibliographystyle{plain}
\bibliography{KDMCEiron,CSC}

\end{document}